\documentclass[12pt]{article}
\usepackage{eurosym}
\usepackage{amssymb,amsmath,epsfig}
\usepackage{color}
\usepackage{amsmath,amsfonts,amssymb}
\usepackage{geometry}
\begin{document}
\title{\textbf{Complexity Factor For Static Anisotropic Self-Gravitating Source in $f(R)$ Gravity}}

\author{G. Abbas \thanks{ghulamabbas@iub.edu.pk}
and H. Nazar \thanks{hammadnazar350@gmail.com}
\\Department of Mathematics The Islamia University\\ of Bahawalpur,
Bahawalpur, Pakistan.}
\date{}
\maketitle
\begin{abstract}
In a recent paper, Herrera \cite{2} (L. Herrera: Phys. Rev. D97, 044010(2018)) have proposed a new definition of complexity for static self-gravitating fluid in General Relativity. In the present article, we implement this definition of complexity for static self-gravitating fluid to case of $f(R)$ gravity.
Here, we found that in the frame of $f(R)$ gravity the definition of complexity proposed by Herrera, entirely based on the quantity known as complexity factor which appears in the orthogonal splitting of the curvature tensor. It has been observed that fluid spheres possessing homogenous energy density profile and isotropic pressure are capable to diminish their the complexity factor. We are interested to see the effects of $f(R)$ term on complexity factor of the self-gravitating object. The gravitating source with inhomogeneous energy density and anisotropic pressure have maximum value of complexity. Further, such fluids may have zero complexity factor if the effects of inhomogeneity in energy density and anisotropic pressure cancel the effects of each other in the presence of $f(R)$ dark source term. Also, we have found some interior exact solutions of modified $f(R)$ field equations satisfying complexity criterium and some applications of this newly concept to the study of structure of compact objects are discussed in detail. It is interesting to note that previous results about the complexity for static self-gravitating fluid in General Relativity can be recovered from our analysis if $f(R)=R$, which General Relativistic limit of $f(R)$ gravity. Some future research directions have been mentioned in the end of the summary.
\end{abstract}


\section{Introduction}

In current scenario General relativity (GR) have discussed numerous critical issues, such as physical
behavior of gravitating source, astrophysical bodies, gravitating physics, interstellar objects, rushing
cosmology, neutron stars and clusters of galaxies. These provide us the key perception to the accelerating
evolution of the cosmos. But at this time, there is need to explain more recent work for the vanishing of complexity
factor in the significance of self-gravitating fluid distribution for relativistic structures. A lot of
discussion about the complexity has been assessed in various fields of science. Now in this attention several researchers have given
systematic work that is shown in \cite{2}-\cite{11}.

Among the various definitions of complexity that have been planed up until now, a large portion of them depend on ideas,
for example information and entropy, and depend on the natural thought that complexity should, somehow,
amount a fundamental property showing the models, existing in the interior framework. Usually, the concept of
complexity in material science paradigms by taking the ideal structure (episodic behavior) and the disengaged
perfect gas, as cases of most straight forward structures and consequently as arrangements with null complexity.
An ideal structure is totally arranged and the atoms are organized after particular principles of symmetry.
The probability distribution for the conditions open to the ideal structure is based on a common condition of ideal
symmetry, in other words it has minimum data fulfilment. On the other hand, the inaccessible perfect
gas is totally scattered. The structure can be produced in any of its open condition by the similar prospect.

Lopez-Ruiz et al. \cite{12} has investigated the concept of instability, which formulate the "distance"
from the equiprobable scattering of the nearby situations of the structure. A lot of renowned astronomers \cite{13}-\cite{16}
have developed the work in complexity for the self-gravitating objects. The notion of self-gravitating structure is
naturally related to the interior fluid distribution, is not associated to instability, somewhat it comes from
the essential presumption that the simplest configuration is shown with the homogeneous matter through ideal pressure. Having
accepted this reality as an intuitive meaning of a terminating complexity structure, the real meaning of obscurity nature will rise
in the advancement of the essential hypothesis of self-gravitating source dense paradigms, in regards of general relativity. Now the
present work suggest to modify theory of gravity for self gravitating fluid distribution in static inhomogeneous region of complexity.
The elementary incentive for this effort exits in complete detail for the state of complexity of the gravitating structure comprised
in \cite{2,8}, \cite{17}-\cite{19}.

Herrera and Santos \cite{20} studied the effects of anisotropy on the evolution of static gravitating source.
Herrera et al. \cite{21} modeled the spherically symmetric gravitating source accompanied with Ricci invariant arrived
from orthogonal spliting of Riemann tensor. Also, Herrera et al. \cite{22} examined the impacts of spherically symmetric dust
on the structures of LTB metrics. Furthermore, Herrera et al. \cite{23} discussed the constancy of shear free
state which depends on the progression equation of the shear tensor and originate that the key factor is performed by Trace-Free tensor $ Y $.
Currently, Herrera and his coworkers \cite{2, 24} conferred through results of self-gravitating paradigms under scalar functions.
Sharif and Zaeem \cite{5} have investigated the imperfect charged dissipative fluid distribution for cylindrically symmetric
self-gravitating source with scalar structure defined by Riemann tensor. Sharif and Yousaf \cite{25} evaluated the consequences
of stable structures in spherically symmetric non-static spacetime under evolution of imperfect fluid.  Herrera et al. \cite{35}
introduced perturbation system in disequilibrium dynamics of gigantic objects and one can determine the disequilibrium conditions
of self-regulating of adiabatic index with considering heat flow transmission inside the interstellar bodies. Chan et al. \cite{36} studied
the spherically symmetric models accompanied by expansion-free perfect fluid source for the degree of gravitational collapse and also
analyzed the results of newtonian effects approaching from degeneracy that rises in the disequilibrium area.

Nojiri and Odintsov \cite{37} took the first initiate to present the notional and thoughtful idea in $f(R)$ theory of gravity for the
behavior of rushing universe. A motivational debate \cite{38, 39} of dark matter regions on the configuration system, many feasible
astral objects discussed in Einstein cosmology and modified theory of gravity $f(R)$. Cembranos et al.\cite{40}
analyzed the impacts of huge-gauge body construction in rushing growth of universe in context of $f(R)$ metric theory and also
studied the gravitating contents in gravitational collapse with non-static inhomogeneous fluid. Santos et al.\cite{41} established the feasibility
conditions originating on general Ricci scalar function $f(R)$. Their technique can be presumed to compel several
probable modified gravity $f(R)$ paradigms under appropriate corporal context. Durrer and Maartens \cite{42} investigated
the noteworthy solutions for celestial structure configuration in background of $f(R)$ paradigms. Cognola et al. \cite{43}
concluded that insights of four major classifications taken from observationally unswerving dark energy (DE) $f(R)$ metric gravity
structures. Kainulainen et al. \cite{44} explored the Tolman-Oppenheimer-Volkoff equation in framework of both metric $f(R)$
and Palatini context with the formation of interior and exterior objects. Fay et al. \cite{45} conferred the cosmological
dynamics of Palatini type $f(R)$ gravity with unlike modified metric $f(R)$ models. Different researchers \cite{46} presented local and cosmological
parameters in distinct $f(R)$ paradigms.

The symmetry of paper trails: In next section, we have established the geometry of the gravitating
anisotropic fluid source, variables related to the spherically symmetric static interior region, the modified Einstein field equations in $f(R)$ background and useful settlements used thoroughly in this paper. The debate on the orthogonal splitting of the curvature tensor and other scalar
functions have been presented in detail in section \textbf{3}. Later section, express the exact results of Einstein field
equation through disappearing of complexity factor. Finally, the last section includes the summary of the work.
\section{Anisotropic Self-gravitating fluid distribution and its interrelated variables}
In this connection to present the physical significance of the gravitating source inside the
celestial object formed by perfect fluid distribution and described with related variables under
$f(R)$ formalism. For this persistence to express the usual Einstein-Hilbert (EH) action in form of General Relativity (GR),
\begin{equation}\label{1}
S_{EH}=\frac{1}{2\kappa}\int d^4x\sqrt{-g}R.
\end{equation}
As for in $f(R)$ context Einstein Hilbert (EH) action can be defined as
\begin{equation}\label{2}
S_{modif}=\frac{1}{2\kappa}\int d^4x\sqrt{-g}(f(R)+L_{(matter)}).
\end{equation}
Here $S_{M}$, $\kappa$, $L_{(matter)}$ and $g$ denotes the modified action source of the generic function of Ricci
scalar $R$, the coupling constant, determine the role of matter contents and the determinant of the metric tensor, respectively.
The following field equations in background of metric $f(R)$ notion are achieved by the variation of equation (\ref{2})
w.r.t. $g_{\alpha\beta}$,
\begin{equation}\label{3}
F(R)R_{\alpha\beta}-\frac{1}{2}f(R)g_{\alpha\beta}-\nabla_{\alpha}\nabla_{\beta}F(R)+g_{\alpha\beta}\nabla^{\alpha}\nabla_{\alpha}F(R)=\kappa T_{\alpha\beta}.
\end{equation}
where $T_{\alpha\beta}$ stands for stress matter energy-momentum tensor, $F(R)=\frac{d f(R)}{d R}$ , $\nabla_{\alpha}$ and $\nabla^\alpha\nabla_\alpha$ are the covariant derivative operator and D'Alembertian, respectively. Above equation (\ref{3}) can be re-manipulated as given under,
\begin{equation}\label{4}
G_{\alpha\beta}=\frac{\kappa}{F}\left(T_{\alpha\beta}^m+T_{\alpha\beta}^D\right),
\end{equation}
where
\begin{equation}\label{5}
T_{\alpha\beta}^D=\frac{1}{\kappa}\left[\frac{f(R)-RF(R)}{2}g_{\alpha\beta}+\nabla_{\alpha}\nabla_{\beta}F(R)-g_{\alpha\beta}\nabla^{\alpha}\nabla_{\alpha}F(R)\right],
\end{equation}
is the effective energy-momentum tensor of the self-gravitating source, bounded by interior region
configuration inside the interstellar model with $f(R)$ modified theory of gravity.
We take an anisotropic matter bounded by spherically symmetric static spacetime.
The interior metric representing the source of self-gravitating fluid is given by
\begin{equation}\label{6}
ds^{2}=e^{\nu}dt^{2}-e^{\lambda}dr^{2}-r^{2}
(d\theta^{2}+\sin^{2}\theta d\phi^{2}),
\end{equation}
where $\nu$ and $\lambda$ are dependable functions of $r$ and coordinates are labeled with
$(x^{0},x^{1},x^{2},x^{3})=(t,r,\theta,\phi)$. Here the components of tensor $T_{\alpha\beta}^m$ play
an important role for the physical mechanism of the object and to evolve, some techniques are in \cite{2, 47}.
In this interpretation, following the Bondi approach \cite{47}, we presumed gently Minkowski coordinates $(\tau,x,y,z)$

$d\tau = e^\frac{\nu}{2}dt; \quad dx = e^{\frac{\lambda}{2}}dr; \quad dy = rd\theta; \quad dz = r sin\theta d\phi$,

Then, under given bar sign shows the energy momentum tensor of the Monkowskian components

$\bar{T^{0}_{0}}={T^{0}_{0}}; \quad\quad \bar{T^{1}_{1}}={T^{1}_{1}}; \quad\quad \bar{T^{2}_{2}}={T^{2}_{2}}; \quad\quad \bar{T^{3}_{3}}={T^{3}_{3}}.$

Now we consider that when seen with proper distance through fluid observer, the $\mu$ is energy density of
fluid source related with space, the $P_{r}$ is radial pressure and $P_{\bot}$ is tangential pressure. Consequently, in the
Minkowski coordinates the matter tensor components are given as

\begin{eqnarray}\nonumber
\left(
  \begin{array}{cccc}
    \mu & 0 & 0 & 0 \\
    0 & P_{r} & 0 & 0 \\
    0 & 0 & P_{\bot} & 0 \\
    0 & 0 & 0 & P_{\bot} \\
  \end{array}
\right).
\end{eqnarray}

Hence
\begin{equation}\label{7}
{T^{0}_{0}}=\bar{T^{0}_{0}}=\mu,
\end{equation}
\begin{equation}\label{8}
{T^{1}_{1}}=\bar{T^{1}_{1}}=-P_{r},
\end{equation}
\begin{equation}\label{9}
{T^{2}_{2}}={T^{3}_{3}}=\bar{T^{2}_{2}}=\bar{T^{3}_{3}}=-P_{\bot}.
\end{equation}
\subsection{The Modified field equations in context of $f(R)$ metric theory}
The relevant field equations with $f(R)$ theory of gravity taken from equation (\ref{3}) reads that
\begin{eqnarray}\label{10}
-\left[-\frac{1}{r^2}+e^{-\lambda}\left(\frac{1}{r^2}-\frac{\lambda'}{r}\right)\right]=\frac{8\pi}{F}\left[\mu+\frac{1}{\kappa}\left\{\frac{f(R)-RF(R)}{2}
+\frac{F''}{e^{\lambda}}+\frac{2F'}{re^{\lambda}}-\frac{\lambda F'}{2e^{\lambda}}\right\}\right],
\end{eqnarray}
\begin{eqnarray}\label{11}
-\left[\frac{1}{r^2}-e^{-\lambda}\left(\frac{1}{r^2}+\frac{\nu'}{r}\right)\right]=\frac{8\pi}{F}\left[P_{r}-\frac{1}{\kappa}\left\{\frac{f(R)-RF(R)}{2}
+\frac{\nu'F'}{2e^{\lambda}}+\frac{2F'}{re^{\lambda}}\right\}\right],
\end{eqnarray}
\begin{eqnarray}\nonumber
&&\frac{e^{-\lambda}}{4}\left[2\nu''+\nu'^2-\lambda'\nu'+2\frac{\nu'-\lambda'}{r}\right]\\&&=\frac{8\pi}{F}\left[P_{\bot}-\frac{1}{\kappa}\left\{\frac{f(R)-RF(R)}{2}
+\frac{F''}{e^{\lambda}}+\frac{\nu' F'}{2e^{\lambda}}-\frac{\lambda' F'}{2e^{\lambda}}+\frac{2F'}{re^{\lambda}}\right\}\right]\label{12}.
\end{eqnarray}

Here $'=\frac{\partial}{\partial r}$. The conservation of modified energy-momentum tensor gives the following equations for the hydrostatic equilibrium
\begin{equation}\label{13}
P'_{r}=-\frac{\nu'(\mu+P_{r})}{2}+\frac{2}{r}(P_{\bot}-P_{r})-D_{1}.
\end{equation}
where $D_{1}$ is the component of dark source term due to $f(R)$ gravity, which is given by
\begin{eqnarray}\nonumber
D_{1}=&&\frac{1}{\kappa}\Big[\left\{\frac{1}{e^{\lambda}}\left(-\frac{f(R)-RF(R)}{2}-\frac{\nu'F'}{2e^{\lambda}}-\frac{2F'}{re^{\lambda}}\right)\right\}_{,1}
+\frac{\nu'}{2e^{2\lambda}}\left\{F''-\frac{\lambda' F'}{2}-\frac{\nu' F'}{2}\right\}\\\nonumber&&
+\frac{\lambda'}{e^{\lambda}}\left\{-\frac{f(R)-RF(R)}{2}-\frac{\nu'F'}{2e^{\lambda}}-\frac{2F'}{re^{\lambda}}\right\}\\&&
\frac{2}{re^{2\lambda}}\left\{F''-\frac{\lambda' F'}{2}\right\}\Big],\label{72}
\end{eqnarray}

This is the generalized Tolman-Oppenheimer-Volkoff equation \cite{44}, for anisotropic fluid in $f(R)$ gravity.

Alternatively, using (\ref{11}), we get
\begin{equation}\label{14}
\nu'=2\frac{(m+\frac{4\pi P_{r}r^3}{F})}{r(r-2m)}-\frac{r^3}{r(r-2m)F}\left\{\frac{f(R)-RF(R)}{2}+\frac{\nu' F'}{2e^{\lambda}}+\frac{2F'}{re^{\lambda}}\right\}.
\end{equation}
Equation (\ref{13}) may be written in the following form
\begin{eqnarray}\nonumber
P'_{r}=&&-\frac{(m+\frac{4\pi P_{r}r^3}{F})}{r(r-2m)}(\mu+P_{r})+\frac{r^2(\mu+P_{r})}{2(r-2m)F}\left\{\frac{f(R)-RF(R)}{2}+\frac{\nu' F'}{2e^{\lambda}}+\frac{2F'}{re^{\lambda}}\right\}\\&&+\frac{2}{r}(P_{\bot}-P_{r})-D_{1}\label{15}.
\end{eqnarray}
The general mass function $m$ is as given below
\begin{equation}\label{16}
R^{3}_{232}=\frac{2m}{r}=1-e^{-\lambda}.
\end{equation}
or,
\begin{equation}\label{17}
m=4\pi\int^{r}_{0}\frac{\tilde{r}^2\mu}{F}d\tilde{r}.
\end{equation}
Here the components of the four-velocity vector are
\begin{equation}\label{18}
v^{\alpha}=(e^{-\frac{\nu}{2}},0,0,0).
\end{equation}
The four-acceleration $a_{\alpha}=v_{\alpha;\beta} v^{\beta}$, the only non-zero
component is
\begin{equation}\label{19}
a_{1}=-\frac{\nu'}{2}.
\end{equation}
From equation (\ref{7})-(\ref{9}), energy-momentum the gravitating fluid is as follows:
\begin{equation}\label{20}
T^{\beta}_{\alpha}=\mu v^{\beta}v_{\alpha}-P h^{\beta}_{\alpha}+\Pi^{\beta}_{\alpha}.
\end{equation}
with
\begin{eqnarray}\nonumber
&&\Pi^{\beta}_{\alpha}=\Pi(\psi^{\beta}\psi_{\alpha}+\frac{1}{3}h^{\beta}_{\alpha}); \quad P=\frac{\tilde{P_{r}}+2P_{\bot}}{3};\\&&
\quad \Pi=P_{r}-P_{\bot}; \quad h^{\beta}_{\alpha}=\delta^{\beta}_{\alpha}-v^{\beta}v_{\alpha},\label{21}
\end{eqnarray}
where non-zero component of $\psi^{\beta}$ is
\begin{equation}\label{22}
\psi^{\beta}=(0,e^{-\frac{\lambda}{2}},0,0),
\end{equation}
and its properties are $\psi^{\beta}v_{\beta}=0, \quad \psi^{\beta}\psi_{\beta}=-1$.
\subsection{The Riemann Curvature and Weyl curvature tensor}
It is convenient to express the Riemann curvature tensor in terms of conformal
curvature tensor $C^{\rho}_{\alpha\beta\mu}$, the Ricci tensor $R_{\alpha\beta}$ and the Ricci
scalar $R$, as
\begin{eqnarray}\nonumber
R^{\rho}_{\alpha\beta\mu}=&&C^{\rho}_{\alpha\beta\mu}+\frac{1}{2}R^{\rho}_{\beta}g_{\alpha\mu}-\frac{1}{2}R_{\alpha\beta}\delta^{\rho}_{\mu}
+\frac{1}{2}R_{\alpha\mu}\delta^{\rho}_{\beta}\\&&-\frac{1}{2}R^{\rho}_{\mu}g_{\alpha\beta}-\frac{1}{6}R(\delta^{\rho}_{\beta}g_{\alpha\mu}
-g_{\alpha\beta}\delta^{\rho}_{\mu}).\label{23}
\end{eqnarray}
In case of spherically symmetric structure, the magnetic part of the conformal curvature tensor becomes identically zero and only it can be defined in form of electric part $(E_{\alpha\beta}=C_{\alpha\gamma\beta\delta}v^{\gamma}v^{\delta})$ as
\begin{eqnarray}
C_{\mu\nu\kappa\lambda}=(g_{\mu\nu\alpha\beta}g_{\kappa\lambda\gamma\delta}-\eta_{\mu\nu\alpha\beta}\eta_{\kappa\lambda\gamma\delta})v^{\alpha}v^{\gamma}E^{\beta\delta}.\label{24}
\end{eqnarray}
where $ g_{\mu\nu\alpha\beta}= g_{\mu\alpha}g_{\nu\beta}-g_{\mu\beta}g_{\nu\alpha}$, and $\eta_{\mu\nu\alpha\beta}$ is the Levi-Civita tensor.
The formula of $E_{\alpha\beta}$ is rewritten as
\begin{equation}\label{25}
E_{\alpha\beta}=E(\psi_{\alpha}\psi_{\beta}+\frac{1}{3}h_{\alpha\beta}),
\end{equation}
with
\begin{equation}\label{26}
E=-\frac{e^{-\lambda}}{4}\left[\nu''+\frac{\nu^{'2}-\lambda'\nu'}{2}-\frac{\nu^{'}-\lambda'}{r}+\frac{2(1-e^{\lambda})}{r^2}\right],
\end{equation}
and fulfill the following conditions:
\begin{equation}\label{27}
E^{\alpha}_{\alpha}=0, \quad E_{\alpha\gamma}=E_{(\alpha\gamma)}, \quad E_{\alpha\gamma}v^{\gamma}=0.
\end{equation}
\subsection{The formulation of Tolman mass and the mass function}
This section, we introduce two definitions of mass for interior of object and their relationship with conformal curvature
tensor. Later, will be used for justification of the complexity factor.

With the help of field equations (\ref{10})-(\ref{12}) and the given mass function (\ref{16}), we get
\begin{equation}\label{28}
m=\frac{4\pi r^3}{3F}(\mu+P_{\bot}-P_{r})+\frac{Er^3}{3}+\frac{r^3}{6F}\left\{\frac{f(R)-RF(R)}{2}+\frac{2F'}{re^{\lambda}}\right\},
\end{equation}
It may be written as
\begin{equation}\label{29}
E=-\frac{4\pi}{r^3}\int^{r}_{0}\frac{\tilde{r}^3}{F}(\mu'-\frac{\mu F'}{F})d\tilde{r}+\frac{4\pi}{F}(P_{r}-P_{\bot})-\frac{1}{2F}\left\{\frac{f(R)-RF(R)}{2}+\frac{2F'}{re^{\lambda}}\right\},
\end{equation}
lastly, using (\ref{29})  in (\ref{28}), one gets
\begin{equation}\label{30}
m(r)=\frac{4\pi r^3\mu}{3F}-\frac{4\pi}{3}\int^{r}_{0}\frac{\tilde{r}^3}{F}(\mu'-\frac{\mu F'}{F})d\tilde{r}.
\end{equation}
The physical significance of the self-gravitating fluid dispersal of (\ref{29}) is basically known by two quantities of
density inhomogeneity and local anisotropy of pressure in $f(R)$ getting through conformal curvature tensor, however in the
instance of a homogeneous mass density distribution, desirable variation tempted through inhomogeneity to describe the mass
function given in equation (\ref{30}).

Another important conformation  was presented by Tolman few decades ago in the explanation of energy source of the matter
surface. For a static distribution of the spherically symmetric object the Tolman mass \cite{2, 48} is given by
\begin{equation}\label{31}
m_{T}=4\pi\int^{r\Sigma}_{0}r^2e^\frac{(\nu+\lambda)}{2}\frac{1}{F}(T^{0}_{0}-T^{1}_{1}-2T^{2}_{2})dr.
\end{equation}
The purpose of Tolman's formula to produce the estimation of the whole mass energy of the structure, through no assurance to its localization.
Now we describe the mass function under $f(R)$ context taken totaly interior of the spherically surface with radius $r$.
\begin{equation}\label{32}
m_{T}=4\pi\int^{r}_{0}\tilde{r}^2e^\frac{(\nu+\lambda)}{2}\frac{1}{F}(T^{0}_{0}-T^{1}_{1}-2T^{2}_{2})d\tilde{r}.
\end{equation}
The noticeable behavior of the "effective inertial mass" as named by $m_{T}$ played with an extension on the universal
theory of energy for the local surface , the detail is given in \cite{9, 19, 20}
\begin{eqnarray}\nonumber
&&m_{T}=e^\frac{(\nu+\lambda)}{2}\left\{m(r)+\frac{4\pi P_{r}r^3}{F}-\frac{r^3}{2F}\left(\frac{f(R)-R F(R)}{2}+\frac{\nu' F'}{2e^{\lambda}}+\frac{2 F'}{re^{\lambda}}\right)\right\}\\&&+\int^{r}_{0}\frac{\tilde{r}^2}{F}e^\frac{(\nu+\lambda)}{2}\left(\frac{f(R)-R F(R)}{2}+\frac{F''}{2e^{\lambda}}
-\frac{\lambda'F'}{4e^{\lambda}}+\frac{3\nu'F'}{4e^{\lambda}}+\frac{2F'}{re^{\lambda}}\right)d\tilde{r}.\label{33}
\end{eqnarray}
By the information of equation (\ref{14}), one can also finds
\begin{eqnarray}\nonumber
&&m_{T}=e^\frac{(\nu-\lambda)}{2}\nu'\frac{r^2}{2}\\&&+\int^{r}_{0}\frac{\tilde{r}^2}{F}e^\frac{(\nu+\lambda)}{2}\left(\frac{f(R)-R F(R)}{2}+\frac{F''}{2e^{\lambda}}-\frac{\lambda'F'}{4e^{\lambda}}+\frac{3\nu'F'}{4e^{\lambda}}+\frac{2F'}{re^{\lambda}}\right)d\tilde{r}.\label{34}
\end{eqnarray}
The overhead expression gives the physical significance of the self-gravitating source of $m_{T}$ is known as "effective inertial mass".
Definitely, used by equation (\ref{19}), is the gravitational acceleration $(a=\psi^{\alpha}a_{\alpha})$ of a test particle, in case of static
gravitational field the test particle instantly at rest followed by \cite{9}
\begin{eqnarray}
a=\frac{e^\frac{-\nu}{2}}{r^2}\left\{m_{T}-\int^{r}_{0}\frac{\tilde{r}^2}{F}e^\frac{(\nu+\lambda)}{2}\left(\frac{f(R)-R F(R)}{2}+\frac{F''}{2e^{\lambda}}-\frac{\lambda'F'}{4e^{\lambda}}+\frac{3\nu'F'}{4e^{\lambda}}+\frac{2F'}{re^{\lambda}}\right)d\tilde{r}\right\}.\label{35}
\end{eqnarray}
The more appealing debate will provide behavior of $m_{T}$ in next section. Differentiate equation (\ref{34})
with respect to $r$ (for this attention slightly calculative work see in more detail \cite{9}), using field
equations and equation (\ref{33}), we have
\begin{eqnarray}\nonumber
&&r m'_{T}-3m_{T}=e^\frac{(\nu+\lambda)}{2}r^3\left[\frac{4\pi}{F}(P_{\bot}-P_{r})-E-\frac{F''}{2Fe^{\lambda}}+\frac{\lambda' F'}{4Fe^{\lambda}}\right]\\&&
+\frac{r^2}{F}e^\frac{(\nu+\lambda)}{2}\left(\frac{f(R)-R F(R)}{2}+\frac{F''}{2e^{\lambda}}-\frac{\lambda'F'}{4e^{\lambda}}+\frac{3\nu'F'}{4e^{\lambda}}+\frac{2F'}{re^{\lambda}}\right).\label{36}
\end{eqnarray}
Here it is easy to formulate the integral form, so that
\begin{eqnarray}\nonumber
 m_{T}=&&(m_{T})_{\Sigma}\left(\frac{r}{r_{\Sigma}}\right)^3-r^3\int^{r_{\Sigma}}_{r}\Big[\frac{e^\frac{(\nu+\lambda)}{2}}{\tilde{r}}\left\{\frac{4\pi}{F}(P_{\bot}-P_{r})-E-\frac{(F''
-\frac{\lambda' F'}{2})}{2Fe^{\lambda}}\right\}\\&&+
\frac{e^\frac{(\nu+\lambda)}{2}}{\tilde{r}^2F}\left(\frac{f(R)-R F(R)}{2}+\frac{F''}{2e^{\lambda}}-\frac{\lambda'F'}{4e^{\lambda}}+\frac{3\nu'F'}{4e^{\lambda}}+\frac{2F'}{\tilde{r}e^{\lambda}}\right)\Big]d\tilde{r}.\label{37}
\end{eqnarray}
Applying equation(\ref{29}), which follows
\begin{eqnarray}\nonumber
 m_{T}=&&(m_{T})_{\Sigma}\left(\frac{r}{r_{\Sigma}}\right)^3-r^3\int^{r_{\Sigma}}_{r}\Big[e^\frac{(\nu+\lambda)}{2}\Big\{\frac{8\pi}{\tilde{r}F}(P_{\bot}-P_{r})
 +\frac{4\pi}{\tilde{r}^4}\int^{\tilde{r}}_{0}\frac{\tilde{r}^3}{F}\left(\mu'-\frac{\mu F'}{F}\right)d\tilde{r}\\\nonumber&&+\frac{1}{2F\tilde{r}}\left(\frac{f(R)-R F(R)}{2}+\frac{2F'}{\tilde{r}e^{\lambda}}\right)-\frac{(F''
-\frac{\lambda' F'}{2})}{2\tilde{r}Fe^{\lambda}}\Big\}\\&&+
\frac{e^\frac{(\nu+\lambda)}{2}}{\tilde{r}^2F}\left(\frac{f(R)-R F(R)}{2}+\frac{F''}{2e^{\lambda}}-\frac{\lambda'F'}{4e^{\lambda}}+\frac{3\nu'F'}{4e^{\lambda}}+\frac{2F'}{\tilde{r}e^{\lambda}}\right)\Big]d\tilde{r}.\label{38}
\end{eqnarray}
Equation (\ref{38}) conferred the noteworthy issues arrived in second integral defines the effects of density inhomogeneity and local anisotropy of the pressure
on the Tolman's mass in the background of $f(R)$ gravity. We would like to mention that all the results reduce to Herrera \cite{2}, when we take $f(R)=R$.
\section{The orthogonal splitting of the curvature tensor}
Bel \cite{49} considered the orthogonal splitting of the
curvature tensor. For that conjecture, we shall use slight changes in notation which are closely
related to \cite{50}.

Now let us familiarize the under mentioned tensors, used by Bel:
\begin{equation}\label{39}
Y_{\alpha\beta}=R_{\alpha\gamma\beta\delta}v^{\gamma}v^{\delta},
\end{equation}
\begin{equation}\label{40}
Z_{\alpha\beta}=^*R_{\alpha\gamma\beta\delta}v^{\gamma}v^{\delta}=\frac{1}{2}\eta_{\alpha\gamma\epsilon\mu}R^{\epsilon\mu}_{\beta\delta}v^{\gamma}v^{\delta},
\end{equation}
\begin{equation}\label{41}
X_{\alpha\beta}=^*R^{*}_{\alpha\gamma\beta\delta}v^{\gamma}v^{\delta}=\frac{1}{2}\eta^{\epsilon\mu}_{\alpha\gamma}R^{*}_{\epsilon\mu\beta\delta}v^{\gamma}v^{\delta}.
\end{equation}

Here the sign $\ast$ representing dual tensor, in other words, $R^{*}_{\alpha\beta\gamma\delta}=\frac{1}{2}\eta_{\epsilon\mu\gamma\delta}R^{\epsilon\mu}_{\alpha\beta}$.

The orthogonal splitting of the curvature tensor are to be written in rewrite form of these tensors
called curvature tensor (see \cite{50}). Though, in substitution for using the explicit form of the
splitting of curvature tensor (equation.(4.6) in \cite{50}), we shall keep as follows in the
general non-static case detail given in \cite{21}.

Equation (\ref{23}) takes the form, by using Einstein field equations
\begin{equation}\label{42}
R^{\alpha\gamma}_{\beta\delta}=C^{\alpha\gamma}_{\beta\delta}+28\pi T^{[\alpha\gamma]}_{[\beta^{\delta} \delta]}+8\pi T\left(\frac{1}{3}\delta^{\alpha\gamma}_{[\beta^{\delta} \delta]}-\delta^{[\alpha\gamma]}_{[\beta^{\delta}{\delta}]}\right).
\end{equation}
Now split the curvature tensor, using (\ref{20}) into (\ref{42}), then we get
\begin{equation}\label{43}
R^{\alpha\gamma}_{\beta\delta}=R^{\alpha\gamma}_{(I)\beta\delta}+R^{\alpha\gamma}_{(II)\beta\delta}+R^{\alpha\gamma}_{(III)\beta\delta}.
\end{equation}
Here
\begin{equation}\label{44}
R^{\alpha\gamma}_{(I)\beta\delta}=16\pi\mu v^{[\alpha\gamma]}v_{[\beta^{\delta}{\delta}]}-28\pi P h^{[\alpha\gamma]}_{[\beta^{\delta} \delta]}
+8\pi(\mu-3P)\left(\frac{1}{3}\delta^{\alpha\gamma}_{[\beta^{\delta} \delta]}-\delta^{[\alpha\gamma]}_{[\beta^{\delta}{\delta}]}\right),
\end{equation}
\begin{equation}\label{45}
R^{\alpha\gamma}_{(II)\beta\delta}=16\pi\Pi^{[\alpha\gamma]}_{[\beta^{\delta} \delta]},
\end{equation}
\begin{equation}\label{46}
R^{\alpha\gamma}_{(III)\beta\delta}=4v^{[\alpha\gamma]}v_{[\beta^E{\delta}]}-\epsilon^{\alpha\gamma}_{\mu}\epsilon_{\beta\delta\nu}E^{\mu\nu}.
\end{equation}
with
\begin{equation}\label{47}
\epsilon_{\alpha\gamma\beta}=v^{\mu}\eta_{\mu\alpha\gamma\beta}, \quad \epsilon_{\alpha\gamma\beta}v^{\beta}=0,
\end{equation}
In interpretation of spherical symmetric the splitting of the curvature tensor due to insertion of the
magnetic part of the conformal curvature tensor $(H_{\alpha\beta}=^{\ast}C_{\alpha\gamma\beta\delta}v^{\gamma}v^{\delta})$.

From the above solutions, we can sort out three definite tensors in the form of the physical variables like
$Y_{\alpha\beta},Z_{\alpha\beta}$ and $X_{\alpha\beta}$ expressed below:
\begin{equation}\label{48}
Y_{\alpha\beta}=\frac{4\pi}{3}(\mu+3P)h_{\alpha\beta}+4\pi\Pi_{\alpha\beta}+E_{\alpha\beta},
\end{equation}
\begin{equation}\label{49}
Z_{\alpha\beta}=0,
\end{equation}
\begin{equation}\label{50}
X_{\alpha\beta}=\frac{8\pi}{3}\mu h_{\alpha\beta}+4\pi\Pi_{\alpha\beta}-E_{\alpha\beta}.
\end{equation}
The overhead expressions denotes the tensors and can describe it in form of structure scalars, the following study takes
an account of scalar functions (see detail in \cite{21}).

Surely, we may state that four structure scalars can attain through $X_{\alpha\beta}$ and $Y_{\alpha\beta}$ tensors and may also
described these tensors with standard notations $X_{T},X_{TF},Y_{T},Y_{TF}$ and last scalar related to the $Z_{\alpha\beta}$
tensor removed in the static case (as shown in detail in \cite{21})

so the structure scalar results are summarized as follows:
\begin{equation}\label{51}
X_{T}=8\pi\mu,
\end{equation}
\begin{equation}\label{52}
X_{TF}=4\pi\Pi-E,
\end{equation}
by using (\ref{29}) and get,
\begin{eqnarray}\nonumber
X_{TF}=&&4\pi(P_{r}-P_{\bot})+\frac{4\pi}{r^3}\int^{r}_{0}\frac{\tilde{r}^3}{F}(\mu'-\frac{\mu F'}{F})d\tilde{r}-\frac{4\pi}{F}(P_{r}-P_{\bot})\\&&+\frac{1}{2F}\left\{\frac{f(R)-RF(R)}{2}+\frac{2F'}{re^{\lambda}}\right\}\label{53},
\end{eqnarray}
\begin{equation}\label{54}
Y_{T}=4\pi(\mu+3P_{r}-2\Pi),
\end{equation}
\begin{equation}\label{55}
Y_{TF}=4\pi\Pi+E,
\end{equation}
equivalently, with the help of (\ref{29})
\begin{eqnarray}\nonumber
Y_{TF}=&&4\pi(P_{r}-P_{\bot})-\frac{4\pi}{r^3}\int^{r}_{0}\frac{\tilde{r}^3}{F}(\mu'-\frac{\mu F'}{F})d\tilde{r}+\frac{4\pi}{F}(P_{r}-P_{\bot})\\&&-\frac{1}{2F}\left\{\frac{f(R)-RF(R)}{2}+\frac{2F'}{re^{\lambda}}\right\}.\label{56}
\end{eqnarray}
The above solutions of $X_{TF}$ and $Y_{TF}$ give the local anisotropy pressure
\begin{equation}\label{57}
X_{TF}+Y_{TF}=8\pi\Pi,
\end{equation}
The solutions of $Y_{T}$ and $Y_{TF}$ arrange the physical meaning, for this instance utilizing (\ref{55}) into (\ref{37}) and we obtain
\begin{eqnarray}\nonumber
 m_{T}=&&(m_{T})_{\Sigma}\left(\frac{r}{r_{\Sigma}}\right)^3+r^3\int^{{r_\Sigma}}_{r}\Big[\frac{e^\frac{(\nu+\lambda)}{2}}{\tilde{r}}\left\{\frac{Y_{TF}}{F}-E(1-\frac{1}{F})
 +\frac{(F''-\frac{\lambda' F'}{2})}{2Fe^{\lambda}}\right\}\\&&-
\frac{e^\frac{(\nu+\lambda)}{2}}{\tilde{r}^2F}\left(\frac{f(R)-R F(R)}{2}+\frac{F''}{2e^{\lambda}}-\frac{\lambda'F'}{4e^{\lambda}}+\frac{3\nu'F'}{4e^{\lambda}}+\frac{2F'}{\tilde{r}e^{\lambda}}\right)\Big]d\tilde{r}.\label{58}
\end{eqnarray}
Analyzing the equation (\ref{58}) by equation (\ref{38}), $Y_{TF}$ explains the impacts of self-gravitating source
of the complexity factor in context of $f(R)$ theory of gravity for the local anisotropy of pressure and density
inhomogeneity on the Tolman mass. On the other hand, $Y_{TF}$ explains how these two expressions alter the calculation
of the Tolman mass with respect to its calculation of perfect fluid homogeneity. It is also worth evoking that $Y_{TF}$
organized with $X_{TF}$, explores the local anisotropy of the matter dispersion.

\section{Matter Distribution with Disappearing Complexity factor}
In order to discuss the system of three ordinary differential equations (\ref{10})-(\ref{12}),
which are modified Einstein equations for static self-gravitating anisotropic fluid contribution
in five unknown functions$(\nu,\lambda,\mu,P_{r},P_{\bot})$ under metric $f(R)$ theory of gravity. Hence, the
off chance is that we implement the condition $Y_{TF}=0$ which might still require one condition while keeping in mind the final
goal to explain the structure.

Therefore from equation (\ref{56}), the disappearing complexity factor condition is
\begin{eqnarray}\nonumber
\Pi=&&\frac{1}{r^3(1+\frac{1}{F})}\int^{r}_{0}\frac{\tilde{r}^3}{F}(\mu'-\frac{\mu F'}{F})d\tilde{r}\\&&+\frac{1}{8\pi(F+\frac{1}{F^2})}\left\{\frac{f(R)-RF(R)}{2}+\frac{2F'}{re^{\lambda}}\right\}.\label{60}
\end{eqnarray}
After taking the expression (\ref{60}), it can be seen that disappearing complexity factor condition
suggest some homogeneous density and pressure isotropy, otherwise inhomogeneous energy density and pressure anisotropy. Likewise, it
ought to be seen that (\ref{60}) might be viewed as a non-local equation of state in $f(R)$ gravity, another few
researchers did work previously in \cite{51} but this note is slightly different. It is interesting to mention that for $f(R)=R$ above equation reduces to Eq.(58) of Herrera \cite{2}.

The preliminary idea to take the attention for those paradigms is the consideration for the
formulation of the metric function $\lambda$ which has the following \cite{52} form
\begin{equation}\label{61}
e^{-\lambda}=1-\alpha r^2+\frac{3K\alpha}{5r^{2}_{\Sigma}}r^{4},
\end{equation}
here $K$ is constant in the interval $(0,1)$ and and $\alpha=\frac{8\pi\mu_{0}}{3}$.

Considering Eqs.(\ref{10}) and (\ref{17}), we get
\begin{eqnarray}\nonumber
\frac{\mu}{F}=&&\mu_{0}\left(1-\frac{K r^2}{r^{2}_{\Sigma}}\right)-\frac{\left(1-\alpha r^2+\frac{3K\alpha}{5r^{2}_{\Sigma}}r^{4}\right)}{\kappa F}
\Big\{\frac{f(R)-R F(R)}{2\left(1-\alpha r^2+\frac{3K\alpha}{5r^{2}_{\Sigma}}r^{4}\right)}\\&&+F''+\frac{2F'}{r}
-\frac{F'\left(2\alpha r-\frac{12 K \alpha r^3}{5 R^{2}_{\Sigma}}\right)}{2\left(1-\alpha r^2+\frac{3K\alpha}{5r^{2}_{\Sigma}}r^{4}\right)}\Big\},\label{62}
\end{eqnarray}
\begin{eqnarray}\nonumber
m(r)=&&\frac{4 \pi\mu_{0}r^3}{3}\left(1-\frac{3K r^2}{5r^{2}_{\Sigma}}\right)-4\pi\int^r_{0}\tilde{r}^2\frac{\left(1-\alpha \tilde{r}^2+\frac{3K\alpha}{5\tilde{r}^2_{\Sigma}}\tilde{r}^4\right)}{\kappa F}
\Big\{\frac{f(R)-R F(R)}{2\left(1-\alpha \tilde{r}^2+\frac{3K\alpha}{5\tilde{r}^2_{\Sigma}}\tilde{r}^4\right)}\\&&+F''+\frac{2F'}{\tilde{r}}
-\frac{F'\left(\alpha \tilde{r}-\frac{6 K \alpha \tilde{r}^3}{5 \tilde{r}^{2}_{\Sigma}}\right)}{\left(1-\alpha \tilde{r}^2+\frac{3K\alpha}{5\tilde{r}^{2}_{\Sigma}}\tilde{r}^{4}\right)}\Big\}d\tilde{r},\label{63}
\end{eqnarray}
introducing, because (\ref{11}) and (\ref{12}) we can get
\begin{eqnarray}\nonumber
\frac{8\pi\Pi(r)}{F}+\frac{1}{r^2}=&&e^{-\lambda}\left[-\frac{\nu''}{2}-\left(\frac{\nu'}{2}\right)^2+\frac{\nu'}{2r}+\frac{1}{r^2}+\frac{\lambda'}{2}\left(\frac{\nu'}{2}+\frac{1}{r}\right)\right]
\\&&+\frac{1}{F}\left(\frac{-F''}{e^{\lambda}}+\frac{\lambda'F'}{2e^{\lambda}}\right),\label{64}
\end{eqnarray}
Letting the variables
\begin{equation}\label{65}
e^{\nu(r)}=e^{\int(2z(r)-\frac{2}{r})dr},
\end{equation}
and
\begin{equation}\label{66}
e^{-\lambda}=y(r).
\end{equation}
From Eq.(\ref{64}), we get the following differential equation
\begin{eqnarray}\label{67}
\left(1+\frac{F'}{F z}\right)y'+y\left[\frac{2z'}{z}+2z-\frac{6}{r}+\frac{4}{r^2z}+\frac{2F''}{F z}\right]=-\frac{2}{z}\left(\frac{8\pi\Pi(r)}{F}+\frac{1}{r^2}\right).
\end{eqnarray}
In this illustration, the overhead solution is seeing to be a Ricatti expression and made with the
assistance of the line element function $\lambda$ held in form of variable $y(r)$ given by (\ref{61}) and also
alongwith the solution of $\Pi$ is obtained from (\ref{60}) and (\ref{62}). The final firm of the solution is providing
$z$ which can be taken through integration. Definitely, it becomes the solution of two functions $z$ and $\Pi$ in context
of gravity $f(R)$, the metric takes the formation \cite{2,53}.
\begin{eqnarray}\nonumber
ds^2=&&-e^{\int\Big(2z(r)-\frac{2}{r}\Big)dr}dt^2\\\nonumber&&+\frac{z^2(r)e^{\int\Big(2z(r)+\frac{4}{r^2z(r)}+\frac{2F''}{F z(r)}\Big)dr}}{r^6\Big(-2\int\frac{z(r)
\Big(F+8\pi \Pi(r) r^2\Big) e^{\int\Big(2z(r)+\frac{4}{r^2z(r)}+\frac{2F''}{F z(r)}\Big)dr}}{r^8F}dr+C\Big)}dr^2\\&&
+r^2d\theta^2+r^2sin^2\theta d\phi^2.\label{68}
\end{eqnarray}
where $C$ is constant of integration.

Next, obtaining for physical variables
\begin{eqnarray}\label{69}
\frac{4\pi P_{r}}{F}=\frac{z(r-2m)+\frac{m}{r}-1}{r^2}+\frac{(1-\frac{2m}{r})}{F}\left(\frac{f(R)-RF(R)}{2(1-\frac{2m}{r})}+(z+\frac{1}{r})F'\right),
\end{eqnarray}
\begin{equation}\label{70}
\frac{4\pi \mu}{F}=\frac{m'}{r^2}-\frac{(1-\frac{2m}{r})}{2F}\left(\frac{f(R)-RF(R)}{2(1-\frac{2m}{r})}+F''+\frac{2F'}{r}-\frac{F'(\frac{m'}{r}-\frac{m}{r^2})}{(1-\frac{2m}{r})}\right),
\end{equation}
and
\begin{eqnarray}\nonumber
\frac{8\pi P_{\bot}}{F}=&&(1-\frac{2m}{r})\left(z'+z^2-\frac{z}{r}+\frac{1}{r^2}\right)+z\left(\frac{m}{r^2}-\frac{m'}{r}\right)\\\nonumber&&
+\frac{1}{F}\Big[\frac{f(R)-RF(R)}{2}+F''(1-\frac{2m}{r})\\&&+(1-\frac{2m}{r})(z+\frac{1}{r})F'+\left(\frac{m}{r^2}-\frac{m'}{r}\right)F'\Big].\label{71}
\end{eqnarray}
Afterwards, fulfills these conditions $\mu>0$ and $\mu>P_{r}, P_{\bot}$, the progression is unvarying for prominent
solutions to remove the complexity in the gravitating system. Next, to evade the nature of singularity for physical
variables on the boundary surface $\sum$, the solution must satisfy the Darmois junction conditions, by taking a vacumm solution solution in the exterior region of the compact object.
\section{Conclusions}

The concept of complexity factor for the static anisotropic gravitating has been present by Herrera \cite{2}. This is very first step towards the understanding of complexity of a gravitating source in general relativity using the Bondi approach. We have found that in the frame of $f(R)$ gravity, also the definition of complexity based on orthogonal splitting of the curvature tensor.
Current work explains the innovative thoughts of the complexity of the system that frame with
static spherically symmetric anisotropic gravitating matter under $f(R)$ gravity. The purpose
of $f(R)$ gravity such a system is to see wether dark energy components lessens the complexity of the structure or enhance it.

 We have investigated $f(R)$ term agrees with an energy density homogeneity and its isotropic pressure to diminish the $Y_{TF}$ scalar function.
Our main objective is to estimate the amount of complexity that appears in the $Y_{TF}$ scalar function with the effects of $f(R)$. Present research discusses such motives that play the vital behavior in sense of applications:
\begin{itemize}
 \item The structure scalar $Y_{TF}$ holds additions from the inhomogeneous energy density
and the native anisotropy of pressure, joined in a systematic order.

 \item The structure scalar $Y_{TF}$ estimates the degree of the "effective inertial mass" in
the departure state for the homogeneous and imperfect matter of the dark gravitating source, given
by the inhomogeneity energy density and the anisotropy of pressure under $f(R)$ formalism. The $Y_{TF}$
includes the impacts of the electric charge in case of charged matter contribution.

 \item The structure scalar also discuss the degeneracy of fluxes alongwith the contributions
of inhomogeneity energy density and native anisotropic pressure in the presence of usual non-static
dissipative gravitating source. In this case to remove the scalar function $Y_{TF}$ is the essential parameter
for the constancy of the shear free case(see in detail \cite{23}).

 \item It ought to be seen that the complexity is much eminent in prior frame, not just departure for
the homogeneous case, perfect fluid, but here two factors display the worth role in equation (\ref{56}) with
$f(R)$ notion and disappear indistinguishably, get additionally for all arrangements where the two factors
in (\ref{56}) cancel each other. The key role of these factors to disappear the complexity of the system
configuration.

 \item It merits telling us that though commitment of anisotropy of pressure to $Y_{TF}$ is natural, the
commitment of inhomogeneous energy density is not.

 \item Consequently, to present the some valuable results that fulfilling the condition for
disappearing of complexity. As suggested previously, the aim was not to furnish paradigms through particular
celestial intrigue, but rather simply show how such paradigms might be attain, with examples.

\item It is very stimulating to note that for $f(R)=R$ Eqs.(14, 15, 28, 29, 30, 33, 35, 38, 37, 53, 56, 58, 62, 67, 68-70) of the present paper reduce to Eqs.(10, 11, 28, 29, 30, 32, 34, 35, 36, 51, 54, 56, 61, 66, 67-69) Herrera Ref.\cite{2}.
\end{itemize}
The final form of the present work is designed for vanishing of complexity factor, in assistance
of exact solutions of $f(R)$ theory of gravity and also would like to confer in future with other modified
concepts, such as $5D$ Gauss-Bonnet gravity, $f(T)$, $f(R, T)$ and Rastall theory.

\end{document}